\documentstyle[epsfig]{aipproc}

\begin{document}
\title{AI Gamma-Ray Burst Classification:
\\ Methodology/Preliminary Results}

\author{Jon Hakkila$^*$, David J. Haglin$^*$, Richard J. Roiger$^*$,
Robert S. Mallozzi$^{\dagger}$, Geoffrey N. Pendleton$^{\dagger}$,
and Charles A. Meegan$^{\ddagger}$}
\address{$^*$Mankato State University, Mankato MN  56002-8400 \\
$^{\dagger}$University of Alabama in Huntsville, Huntsville, AL 35899 \\
$^{\ddagger}$NASA/Marshall Space Flight Center, Huntsville, AL 35824}

\maketitle

\begin{abstract}
Artificial intelligence (AI) classifiers can be used 
to classify unknowns, refine existing classification parameters, and 
identify/screen out ineffectual parameters. We present an 
AI methodology for classifying gamma-ray bursts, along with some
preliminary results.
\end{abstract}

\section*{Methodology}
Gamma-ray burst (GRB) subclassification is difficult due to complex
burst spectral and temporal behaviors \cite{Dyson96}. Few long-standing 
classification attributes have been known except for those 
identified on the basis of duration and spectral hardness \cite{Kouveliotou93}.
Recent evidence suggests that
(a) bursts can contain high-energy peaks (with emission above 300 keV),
non-high-energy peaks \cite{Pendleton97} or both,
(b) excessive low-luminosity GRB emission is correlated with spectral
hardness and duration for a sample of long, bright bursts \cite{Horack97},
(c) the spectral break energy vs. intensity relation \cite{Liang96} might
indicate that GRB pulse decay is governed by radiative cooling, and 
(d) significant emission below 10 keV exists in 10-15\% of bursts
\cite{Preece96}.

We are developing a tool to automate GRB classification using the
AI technique of knowledge discovery in databases (KDD). This technique
has already been used successfully in optical/infrared astronomy 
\cite{Weir95}: a 300\% increase in survey classification was obtained,
even though the AI classifier used in this studey found only eight of 
40 input attributes to be important.

KDD is usually seen as a four step process: data selection, data pre-processing
and transformation, data mining, and interpretation/evaluation.

{\bf Step 1: Data Selection.}
The choice of attributes can affect the outcome of classification
analysis, since AI classifiers cannot make informed decisions with insufficient
information. Moreover, few GRB attributes are useful in a raw (unprocessed)
format. Although preprocessed data is initially being drawn from the BATSE
database, future expansion to other databases is planned.

{\bf Step 2: Data Pre-Processing and Transformation.}
Table \ref{table} indicates some preprocessed attributes that can be used 
as AI classifier input. These range from single elements, to arrays, and to more
complex data structures. This list is not exhaustive, but merely indicates the
size of the database that can be introduced for classification purposes.

\begin{table}
\caption{Some Potentially Important Classification Data Available 
For Each Burst.}
\label{table}
\begin{tabular}{ddd}
duration & fluence & hardness ratios (e.g. HR21, HR32, HR43) \\
peak flux & location & burst duration from time range of peaks \\
number of peaks & & spectral indices in different energy ranges \\ 
ILF power-law index & & distribution of peaks by spectral hardness \\
spectral break energy & & low-energy flux (below 10 keV) \\
peak fluxes in different energy bands & & spectral evolution summary \\
\end{tabular}
\end{table}

{\bf Step 3: Data Mining.}
Data Mining is the application of one or
more pattern identification algorithms to a specific data set. It
produces a classification structure representative of the
concept classes identified; these are used to document
relationships, verify previous knowledge, and predict future outcome. 
Unknown instances are classified by using the classification structure with an 
appropriate interpretive
algorithm.  Classifiers are supervised (decision trees 
\cite{Quinlan86} and rule sets are trained with known classification
instances), unsupervised (concept hierarchies 
\cite{Gennari89} require learning to be performed without training examples), 
or both (neural networks \cite{Jain96}).

{\bf Step 4: Interpretation/Evaluation.}
We have modified existing KDD techniques by combining them with
the scientific method. By using this approach, we are attempting to
address errors (such as statistical errors and instrumental biases)
leading to improperly-identified subclasses/substructures.

Unsupervised learning is first used to determine if well-defined burst classes
can be discovered. Several unsupervised classifiers are used for comparative 
analysis. After an unsupervised classifier identifies a concept class, the class
instances are presented to supervised learning models so that classification 
structures can be identified depicting the named concepts.

Next, classification success is evaluated. 
Unsupervised
classification methodology relies on internal checks such as inter-class
difference checks, intra-class prototype similarity checks, and 
instance-by-instance classification comparisons.
Success is evaluated in a supervised environment by a 
variety of ``goodness-of-fit''
parameters including predictiveness scores (statistical pull of an attribute
relative to the class mean value), classification correctness (number of
correctly classified instances relative to total number of set instances),
and attribute correlations (to eliminate related attributes). 

The subclasses and/or classification substructures identified are 
carefully studied to determine the extent to which they can be attributed
to instrumental effects and/or observational biases. We rely on our
expertise in working with GRB data and on our use of datasets obtained from
a variety of GRB experiments to identify these biases.

When the data have been corrected for biases, the process begins anew from
the point of unsupervised classification. We believe that this process
will allow us to successfully identify properties of gamma-ray burst classes,
and to optimize differentiation between known or suspected burst subclasses.

\section*{Preliminary Results}

We have begun applying the KDD process to the BATSE 4B Catalog. We have
constructed a test dataset of 954 bursts for which we have chosen six basic
preprocessed attributes: T90 duration, 1024 ms peak flux, channel 2+3 fluence, 
and three hardness ratios (HR21, HR32, and HR43).

\begin{figure}[b!] 
\centerline{\psfig{file=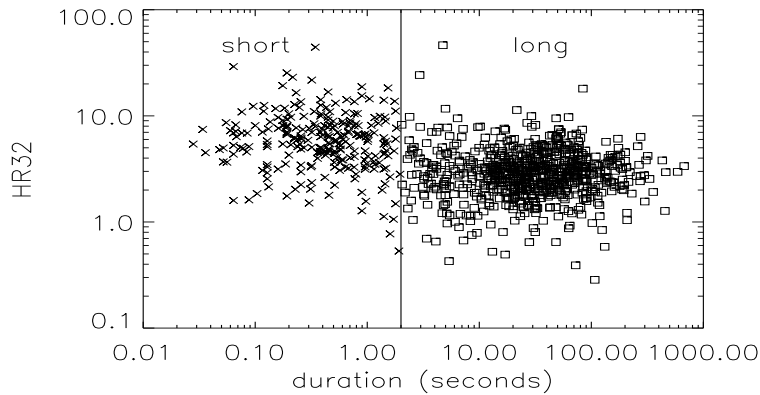,height=2.0in,width=3.5in}}
\vspace{12pt}
\caption{HR32 vs. duration diagram. Long and short bursts train the 
supervised classifier C4.5, even though there is not a clean subclass 
separation.}
\label{fig1}
\end{figure}

The duration bimodality (short bursts have durations $< 2$ seconds; long ones
have durations $\geq 2$ seconds) is the one natural division expected in this
dataset, producing two corresponding ``subclasses'' of slightly differing HR32
hardness ratios (e.g. \cite{Kouveliotou93}). It should be noted that a clean
division does not exist between these subclasses; there is considerable overlap
in the distributions (a histogram of HR32 does not produce strong evidence of
two subclasses; \cite{Smith96}). We have included as Figure \ref{fig1} a 
plot of HR32 vs. duration, so that the reader might verify the efficacy of 
this classification. 

Initially, we used the unsupervised classifier CLASSIT to identify subclasses.
Surprisingly, CLASSIT did not create concept clusters that clearly
defined ``long'' and ``short'' subclasses.
{\it The classifier did not ignore duration and hardness in making its
decision; rather it concluded that there were more statistically significant
subsets in the data than this particular one.} We are currently investigating
this result in greater detail.

We trained the benchmark supervised learning program C4.5 with 25 short
and 25 long bursts. C4.5 built a decision tree to be used for 
classifying new instances of unknown duration, as well as a set of rules
representing decision tree paths. The rules were used to classify the
remaining 904 bursts; these were 
correctly classified as long or short 89.7\% of the time.

We concluded that duration information was being included in the
fluence attribute, and subsequently retrained the classifier after having
removed the fluence attribute.
{\it The classifier still correctly
classified the unknowns 77.2\% of the time without any
duration information.} The rule set is simply:

\begin{center}IF (HR32 $> 4.60$ OR HR21 $> 2.23$) THEN SHORT ELSE LONG
\end{center}

\begin{figure}[b!] 
\centerline{\epsfig{file=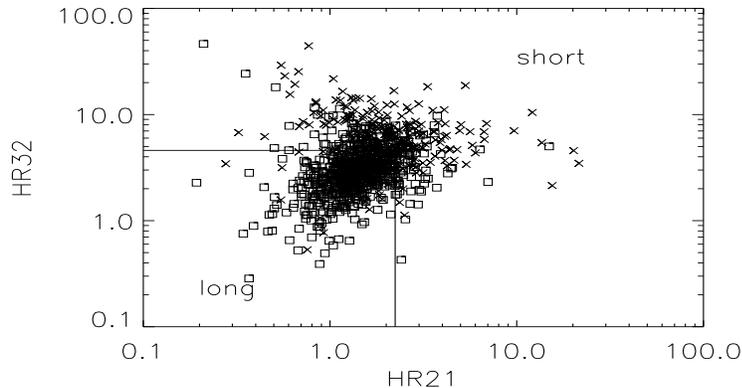,height=2.0in,width=3.5in}}
\vspace{12pt}
\caption{HR32 vs. HR21 plot identifying long and short bursts, 
according to a simple decision tree identified by C4.5 trained on 50 BATSE 
bursts. The rule set identified allows 77\% correct classification of unknown
instances, is independent of peak flux and HR43 (which were ignored by the 
classifier despite the known hardness-intensity correlation), and does not 
include correlations with duration and fluence (not included for analysis).}
\label{fig2}
\end{figure}

Figure \ref{fig2} demonstrates the effect of this simple rule on bursts
with ``unknown'' duration classes. From this rule, C4.5 classified the 681 long 
bursts correctly 77\% of the time and the 223 short bursts 78\% of the time.

We considered the possibility that relative numbers of long vs. short events
might contribute to classification success: in other words, if the knowledge
that 75\% of the bursts are long and 25\% are short was used, then a 75\% 
accuracy could be obtained by guessing that all bursts would be long. To test 
this, we applied the decision tree to a sample of 223 short 
bursts (50\%) and 223 long ones (50\%). The classification accuracy was still 
80.5\% (78\% accuracy for long bursts and 83\% accuracy for long ones), 
although guessing that all bursts would be long would now only produce a 50\% 
accuracy.

The C4.5 results are consistent with those obtained using other supervised
classifiers. We subsequently verified the relative importances of the fluence, 
HR32, and HR21 attributes using the supervised classifier SX-WEB and
Bayesian-based Discriminant Analysis.

\section*{Conclusions and Future Study}

The long and short burst substructures \cite{Kouveliotou93} do
not represent optimum data subclasses, as suggested by inspection of Figure 1
and by results from the unsupervised classifier CLASSIT. Nonetheless, our
analysis indicates strong evidence to support the existence of burst groupings
that relate hardness to duration. It is quite possible that these groups would
be made more distinct by the aid of additional preprocessed attributes.

The two duration groups have distinct fluence differences and correlated 
hardnesses, implying that ensemble analyses examining duration and/or hardness 
for other purposes might not succeed unless these groups are considered 
separately. Analyses that might be affected include the use of durations to 
identify cosmological time dilation \cite{Norris94} \cite{Mitrofanov96}, the 
use of hardness variations to determine cosmological energy shifts, and
the use of fluences (S) in $\log(N>S)$ vs. $\log(S)$ to determine the 
cosmological distance scale \cite{Bloom96} \cite{Lee96}.

These preliminary results demonstrate the power
of applying the KDD process to gamma-ray bursts. The
process has been used here to verify the predictive power of an existing 
subclassification structure;
we will apply KDD in the future to verify other subclasses and/or
classification substructures and to search for previously unknown ones.


\begin{references}
\bibitem{Bloom96}Bloom, J. S., Fenimore, E. E., \& in't Zand, 1996, in 
{\it Gamma-Ray Bursts}, ed. C. Kouveliotou, M. S. Briggs, \& G. J. Fishman, 
(AIP: New York), p. 321.
\bibitem{Dyson96}Dyson, S. E., \& Schaefer, B. E., 1996, 
in {\it Gamma-Ray Bursts}, ed. C. Kouveliotou, M. S. Briggs, \& G. J. Fishman, 
(AIP: New York), p. 363.
\bibitem{Gennari89}Gennari, J. H., Langley, P. \& Fisher,D., 1989, 
{\it Artificial Intelligence}, {\bf 40}, 11.
\bibitem{Horack97}Horack, J. M. \& Hakkila, J. 1997, {\it ApJ}, {\bf 479}, 371.
\bibitem{Jain96}Jain, A.K., Mao, J., \& Mohiuddin, K. M., 1996, 
{\it IEEE Computer}, {\bf 29:3},  31.
\bibitem{Kouveliotou93}Kouveliotou, C., et al., 1993, {\it ApJ}, {\bf 413}, 
L101.
\bibitem{Lee96}Lee, T. T. \& Petrosian, V. 1996, in {\it Gamma-Ray Bursts}, 
ed. C. Kouveliotou, M. S. Briggs, \& G. J. Fishman, (AIP: New York), p. 462.
\bibitem{Liang96}Liang, E. P. \& Kargatis, V. 1996,  in {\it Gamma-Ray Bursts}, 
ed. C. Kouveliotou, M. S. Briggs, \& G. J. Fishman, (AIP: New York), p.  202.
\bibitem{Mitrofanov96}Mitrofanov, I. G., et al., 1996, {\it ApJ}, {\bf 459}, 
570.
\bibitem{Norris94}Norris, J. P. et al., 1994, {\it ApJ},  {\bf 424}, 540.
\bibitem{Pendleton97}Pendleton, G. N., et al., 1997, {\it ApJ} (in press).
\bibitem{Preece96}Preece, R. D., et al., 1996, in {\it Gamma-Ray Bursts}, 
ed. C. Kouveliotou, M. S. Briggs, \& G. J. Fishman, (AIP: New York),  p. 238.
\bibitem{Quinlan86}Quinlan, J. R., 1986, {\it Machine Learning}, {\bf 1}, 81.
\bibitem{Smith96}Smith, I. A. et al. 1996, in {\it Gamma-Ray Bursts}, ed. C. 
Kouveliotou, M. S. Briggs, \& G. J. Fishman, (AIP: New York), p. 102.
\bibitem{Weir95}Weir, N., Fayyad, U.M., \& Djorgovski, S.G., 1995, {\it AJ}, 
{\bf 109}, 2401.
\end{references}
\end{document}